\documentclass[aps,prb,reprint,superscriptaddress]{revtex4-2}

\usepackage{graphicx}
\usepackage{dcolumn}
\usepackage{bm}
\usepackage{amsmath}
\usepackage{amssymb}
\usepackage{mathtools}
\usepackage{lipsum}

\usepackage[utf8]{inputenc}
\usepackage{etoolbox}
\usepackage{soul}
\usepackage[colorlinks=true,citecolor=blue,urlcolor=blue]{hyperref}
\usepackage{tcolorbox}
\usepackage{siunitx}
\usepackage{braket}
\usepackage{todonotes}
\usepackage{array}
\usepackage{multirow}
\usepackage{float}
\usepackage{pifont}
\usepackage{lipsum}
\usepackage{mhchem}

\usepackage{newtxtext,newtxmath}
\DeclareMathAlphabet\mathbfcal{OMS}{cmsy}{b}{n}
\DeclareSymbolFontAlphabet{\mathcal}{symbols}
\DeclareMathAlphabet{\mathcal}{OMS}{cmsy}{m}{n}
\SetMathAlphabet{\mathcal}{bold}{OMS}{cmsy}{b}{n}

\newcommand{\iu}{\mathrm{i}}
\newcommand{\ec}{\mathrm{e}}

\renewcommand{\vec}[1]{\boldsymbol{#1}}
\allowdisplaybreaks

\definecolor{darkpastelgreen}{rgb}{0.01, 0.75, 0.24}

\newcommand{\vmomentum}{\vec{k}}
\newcommand{\mmomentum}{k}
\newcommand{\vmomentumtransfer}{\vec{q}}
\newcommand{\mmomentumtransfer}{q}
\newcommand{\vMomentum}{\vec{K}}

\newcommand{\neutralSemicondState}{\mathrm{DS}}

\begin{document}

\title[Unconventional excitonic insulators in two-dimensional topological materials]{Unconventional excitonic insulators in two-dimensional topological materials}
\author{L. Maisel Licerán}
\email{l.maiselliceran@uu.nl}
\affiliation{Institute for Theoretical Physics and Center for Extreme Matter and Emergent Phenomena, Utrecht University, Princetonplein 5, 3584 CC Utrecht, The Netherlands}
\author{H. T. C. Stoof}
\affiliation{Institute for Theoretical Physics and Center for Extreme Matter and Emergent Phenomena, Utrecht University, Princetonplein 5, 3584 CC Utrecht, The Netherlands}

\date{\today}

\begin{abstract}
    \noindent
	Bound electron-hole pairs in semiconductors known as excitons can form a coherent state at low temperatures akin to a BCS condensate.
	The resulting phase is known as the excitonic insulator and has superfluid properties.
	Here we theoretically study the excitonic insulator in a pair of recently proposed two-dimensional candidate materials with nontrivial band topology.
	Contrary to previous works, we include interaction channels that violate the individual electron and hole number conservations.
	These are on equal footing with the number-conserving processes due to the substantial overlap of Wannier orbitals of different bands, which cannot be exponentially localized due to the nontrivial Chern numbers of the latter.
	Their inclusion is crucial to determine the symmetry of the electron-hole pairing, and by performing mean-field calculations at nonzero temperatures we find that the order parameter in these systems is a chiral $d$-wave.
	We discuss the nontrivial topology of this unconventional state and discuss some properties of the associated Berezinskii-Kosterlitz-Thouless transition.
	In particular, we argue that here it becomes a smooth crossover and estimate the associated temperature to lie between $\SI{50}{\kelvin}$ and $\SI{75}{\kelvin}$ on realistic substrates, over an order of magnitude larger than in the number-conserving approximation where $s$-wave pairing is favored.
	Our results highlight the interplay between topology at the single-particle level and long-range interactions, motivating further research in systems where both phenomena coexist.
\end{abstract}

\maketitle

\section{Introduction}

\noindent
Collective phenomena in many-body systems give rise to surprising material properties and new phases of matter.
Prime examples are superconductivity and superfluidity, which allow for the dissipationless transport of charge and mass, respectively \cite{bardeen1957theory,tinkham2004introduction,annett2004superconductivity,leggett2006quantum,volovik2003universe}.
These phenomena have long been observed as a result of the condensation of Cooper pairs, bound states of two electrons occurring in metals at very low temperatures.
In semiconductors, fermion pairing can also occur at higher temperatures in the form of excitons, which are bound electron-hole pairs.
A coherent state of excitons can form when the binding energy of these pairs exceeds the band gap energy of the semiconductor.
This phase is called the excitonic insulator (EI) and could host exotic phenomena such as superfluidity, anomalous electrical responses and quantum oscillations, and topology, to name a few \cite{jerome1967excitonic,fogler2014high,wang2019prediction,sun2021second,hu2022quantum,wang2023excitonic,shao2024electrical,shao2024quantum,allocca2024fluctuation,yang2024spin,yang2024band,xie2024theory,dong2025topological,nguyen2025perfect,qi2025perfect}.

Recent theoretical and experimental work on exciton physics has mostly focused on two-dimensional (2D) systems \cite{high2012spontaneous,du2017evidence,li2017excitonic,wang2019evidence,ma2021strongly,gu2022dipolar,jia2022evidence,huang2024evidence,zhang2024spontaneous}, which feature reduced dielectric screening of the Coulomb interaction and allow for the suppression of the radiative recombination rate via the use of spatially separated electron and hole layers.
The latter is advantageous for exploring optical properties and exciton Bose-Einstein condensates, but comes at the cost of reduced exciton binding energies due to the weaker interlayer Coulomb interaction.
However, the EI is a BCS-like state which appears due to an instability of the Fermi surface against spontaneous exciton formation, requiring relatively large Coulomb interactions so that the binding energies exceed the gap.
The resulting phase is naturally more robust against recombination processes, and pairing with nonzero angular momentum can further increase this robustness as the correlated particles tend to avoid each other.
For these reasons it is important to understand in detail the EI physics in monolayer systems, where the condition of strong Coulomb interactions can be realized more easily.
As shown here, their behavior can differ greatly from that in bilayer systems, with non-$s$-wave condensation realized in the presence of topology.

\begin{figure}[!t]
    \centering
    \includegraphics[width=\linewidth]{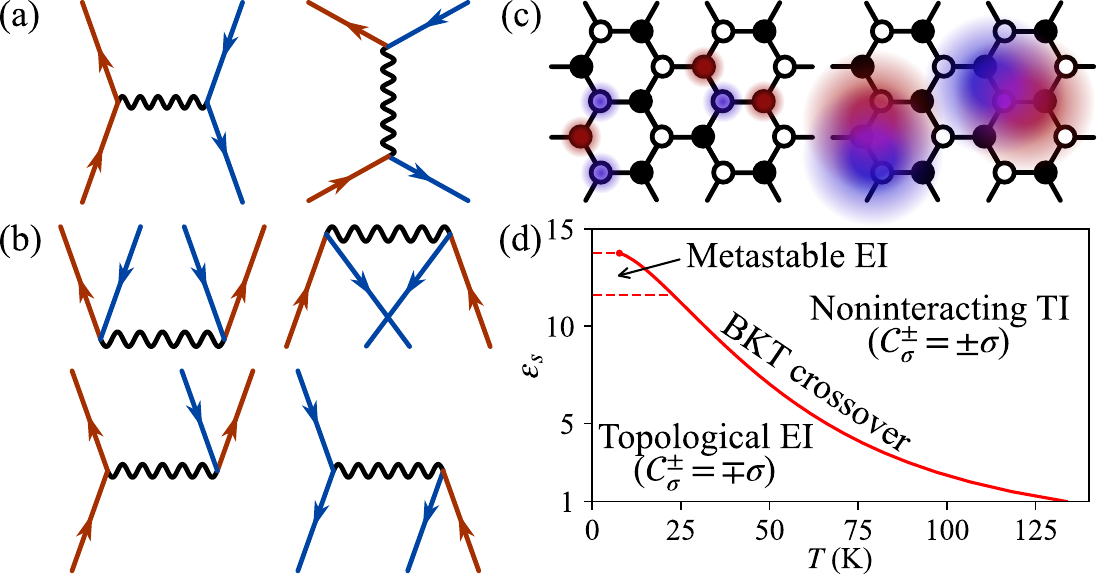}
    \caption{\textbf{(a)} Examples of $\mathrm{U}(1)$-symmetric scattering channels, where the number of ingoing electrons (red lines) is equal to that of outgoing ones, and similarly for the holes (blue lines).
    \textbf{(b)} Examples of $\mathrm{U}(1)$-breaking scattering processes which do not conserve the individual electron and hole numbers.
    \textbf{(c)} Sketch of Wannier orbitals in the case of trivial bands (left) and bands with a nonzero Chern number (right).
    Due to the nonlocalizability of the latter, orbitals corresponding to the conduction and valence bands have significant mutual overlaps in coordinate space and the processes depicted in (b) become relevant.
    \textbf{(d)} Phase diagram of MTCO in the $\varepsilon_{s}$--$T$ plane, where $\varepsilon_{s}$ is the dielectric constant of the material substrate.
    At low $\varepsilon_{s}$ and $T$ we find a topological $d + \iu d$ singlet EI whose Chern numbers for the upper ($+$) and lower ($-$) band with spin $\sigma$ are opposite to those of the underlying noninteracting TI at high $\varepsilon_{s}$ or $T$.
    This state shares the same symmetries as the underlying TI due to the explicit $\mathrm{U}(1)$-symmetry breaking caused by the diagrams in (b).
    As a result, the BKT transition calculated here via the standard theory becomes only a smooth crossover and is accompanied by a gap closing in its vicinity which connects the two topologically distinct phases.
    }
    \label{fig:figIntro}
\end{figure}

We consider the EI in topological monolayer materials with a band inversion accompanied by a phase winding of the electron and hole states around a high-symmetry point \cite{bernevig2006quantum,zhang2009topological,zhu2012band}.
Because the bands in such systems possess a nonzero Chern number, the associated Wannier orbitals are weakly localized \cite{thouless1984wannier,brouder2007exponential}.
When considering Coulomb interactions, this feature manifests in the appearance of scattering processes where the involved quasiparticles can switch bands at the electrostatic vertex, thus breaking the individual electron and hole $\mathrm{U}(1)$ symmetries.
In Fig.\ \hyperref[fig:figIntro]{1b} we show some examples of these $\mathrm{U}(1)$-breaking channels.
Furthermore, the phase winding of the single-particle wave functions causes these channels to couple together different angular momenta, in turn influencing the pairing symmetry of the electron-hole order parameter.

In this work we consider the topological 2D materials \ce{AsO} and \ce{Mo2TiC2O2} (MTCO) introduced in Ref.~\cite{yang2024spin}, which feature such a band inversion around the $\Gamma$ point and whose conduction and valence bands closest to the Fermi surface have unit Chern numbers \footnote{MTCO is a member of the family of MXenes, 2D materials whose prospect for topological and highly correlated phenomena has generated increasing interest in recent years \cite{gogotsi2019rise,khazaei2019recent,salim2019introduction,naguib2021ten,vahidmohammadi2021world,dong2023robust,kumar2023oxygen,kumar2024excitons}}.
In these systems, these two bands have equal parity at the $\Gamma$ point, which inhibits dipole transitions between them.
As a result, the screening of the Coulomb interaction is largely independent of the magnitude of the band gap, which allows for exciton binding energies larger than the gap \cite{jiang2018realizing}.
Ref.~\cite{yang2024spin} only included the $\mathrm{U}(1)$-symmetric channels of Fig.\ \hyperref[fig:figIntro]{1a} on top of the noninteracting bands, where the character of a particle (electron or hole) is conserved at the electrostatic vertex, an approach which is in fact ubiquitous \cite{wang2019prediction,ataei2021evidence,wang2023breakdown,davari2024optical,xie2024theory,yang2024spin}.
Here we show that including all additional $\mathrm{U}(1)$-breaking channels leads to a topological $d + \iu d$ spin-singlet EI, whereas the number-conserving approximation gives an $s$-wave order parameter in the triplet channel.

An important tuning parameter in our calculations is the dielectric constant $\varepsilon_{s}$ of the substrate below the material sample.
Above a certain material-dependent critical value, the dielectric screening of the electron-hole interaction is too strong to sustain an EI and we obtain the band structure of the underlying noninteracting semiconductor, which we refer to as the topological insulator (TI) phase.
Beyond the mean-field level on which we focus in this article, the transition between both is expected to take place via the Berezinskii-Kosterlitz-Thouless mechanism, where bound vortex-antivortex pairs become unbound at a certain critical temperature \cite{kosterlitz1973ordering,nelson1977universal,ryzhov2017berezinskii}.
However, the TI and the EI share the same symmetries due to the inevitable presence of the $\mathrm{U}(1)$-breaking terms.
Accordingly, the BKT mechanism will now lead to a smooth crossover instead of a true phase transition.
In Fig.\ \hyperref[fig:figIntro]{1d} we show the phase diagram of MTCO in the $\varepsilon_{s}$--$T$ plane.
We have estimated that the EI in both materials remains robust up to temperatures as high as $\SI{75}{\kelvin}$ on realistic substrates.

\section{Low-energy model}

\noindent
The materials under consideration are described by a low-energy Hamiltonian $H_{0}(\vmomentum) = \operatorname{diag}[h_{+}(\vmomentum), h_{-}(\vmomentum)]$ that correctly reproduces the $GW$ band structure around the $\Gamma$ point \cite{yang2024spin}. Here,
\begin{equation}
\label{eq:intro:singleParticleH}
	h_{\sigma}(\vmomentum) = \begin{bmatrix}
		h_{0}(\mmomentum) + \sigma \lambda & \alpha \mmomentum_{-}^{2} \\ \alpha \mmomentum_{+}^{2} & h_{0}(\mmomentum) - \sigma \lambda
	\end{bmatrix} \! ,
\end{equation}
with $\mmomentum_{\pm} = \mmomentum_{x} \pm \iu \mmomentum_{y}$, $\mmomentum = |\vmomentum|$, and $h_{0}(\mmomentum) = \epsilon_{0} + c \mmomentum^{2}$.
The diagonal blocks of $H_{0}(\vmomentum)$ describe spin-up and spin-down electrons ($\sigma = \pm$, respectively) in a basis of $p_{x} \pm \iu p_{y}$ ($d_{xy} \pm \iu d_{x^{2} - y^{2}}$) orbitals for \ce{AsO} (MTCO).
The bands of $h_{\sigma}$ have a gap of magnitude $2 \lambda$ at $k = 0$ and nontrivial Chern numbers $\mathcal{C}^{c,v}_{\sigma} = \pm \sigma \operatorname{sgn} \lambda$, with opposite signs for each type of band in the two spin subspaces due to the time-reversal symmetry.

On top of this single-particle picture we consider the repulsive Coulomb interaction $\hat{V} = \frac{1}{2 \mathcal{A}} \sum_{\vec{q}} \hat{n}_{\vmomentumtransfer} V(\vmomentumtransfer) \hat{n}_{{-} \vmomentumtransfer}$, with $\hat{n}_{\vec{q}}$ the Fourier-transformed density operator in the orbital-spin basis corresponding to the $\Gamma$-point eigenstates and $\mathcal{A}$ the surface area of the system.
In App.\ \ref{app:CoulombIntkDotpModel} we show on very general grounds that this form of the interaction must be written in our basis states at the $\Gamma$ point, which in turn leads to the $\mathrm{U}(1)$-breaking terms.
Due to the 2D geometry we neglect screening of the Coulomb interaction, but we do include polarization and substrate effects by taking $V(\vmomentumtransfer) = 2 \pi / [\mmomentumtransfer (\varepsilon + 2 \pi \alpha_{\text{2D}} \mmomentumtransfer)]$.
Here $\alpha_{\text{2D}}$ is the 2D polarizability of the medium, and $\varepsilon = (1 + \varepsilon_{s}) / 2$ is the average between the dielectric constant $\varepsilon_{s}$ of the substrate below the sample and the vacuum above.
We also exclude electron-phonon coupling because upon integrating out the incoherent phonon bath this leads to $\mathrm{U}(1)$-breaking terms like the ones we already consider \cite{zenker2014fate}.
Hence, at least in the static limit, such a coupling will at most lead to a channel-dependent $\varepsilon$ that leaves the qualitative picture unchanged.
We note that Ref.\ \cite{zenker2014fate} additionally considers a coherent phonon mode condensing with the excitons, but such a situation does not apply to our article since the single-particle model we start with is not a semimetal but a semiconductor with a relatively large gap compared with the typical expected phonon energies.

The numerical values of the model parameters in atomic units \footnote{In atomic units, the unit of length is given by the Bohr radius, $a_{0} = \SI{0.529177}{\AA}$, and the unit of energy is the Hartree energy, $E_{\mathrm{h}} = \SI{27.211386}{\electronvolt}$. In particular, the momenta are expressed in units of $a_{0}^{-1}$.} are $\lambda = 0.0030$, $c = 0.198$, $\alpha = -0.534$, $\alpha_{\mathrm{2D}} = 38.7$ for \ce{AsO}, and $\lambda = 0.0021$, $c = -1.251$, $\alpha = 1.602$, $\alpha_{\mathrm{2D}} = 19.7$ for MTCO.

\section{$\boldsymbol{\mathrm{U}(1)}$-breaking terms in the TI phase}

\noindent
Before addressing their consequences on the EI, we briefly explore the effect of the $\mathrm{U}(1)$-breaking channels on excitons in the TI state.
In this context, neglecting them is known as the Tamm-Dancoff approximation (TDA) for the excitonic Bethe-Salpeter equation (BSE).
The TDA drops the $\mathrm{U}(1)$-breaking coupling between resonant and antiresonant processes in the calculation of exciton spectra and keeps only $\mathrm{U}(1)$-symmetric processes \cite{hirata1999time,onida2002electronic,gruning2009exciton,sander2015beyond,maggio2016correlation,sander2017macroscopic}.
This is appropriate when the resonant and antiresonant subspaces are far away in energy, which in particular holds when the exciton binding energies $\epsilon_{b}$ are much smaller than the semiconductor band gap $E_{g}$.
In this case, $\kappa^{2} \epsilon_{b} / 2E_{g} \ll 1$, with $\kappa$ a characteristic overlap between the valence and conduction wave functions.
However, if the binding energy of excitons becomes comparable to the gap, the resonant-antiresonant coupling becomes important.
For this reason we expect the $\mathrm{U}(1)$-breaking channels to influence the onset of the EI phase, and the question arises whether they also play a role on the properties of the EI at low temperatures.

The full BSE goes beyond the variational exciton state $\vert X_{\Phi} \rangle = \sum_{cv \vmomentum} \Phi^{cv}_{\vmomentum} \hat{\psi}^{\dagger}_{c \vmomentum} \hat{\psi}_{v \vmomentum} \vert \neutralSemicondState \rangle$, which automatically leads to the TDA for the envelope wave function $\Phi$.
Here, $c$ ($v$) represents an arbitrary conduction (valence) band with corresponding creation and annihilation operators $\hat{\psi}^{\dagger}_{c(v)\vmomentum}$ and $\hat{\psi}_{c(v)\vmomentum}$, respectively, and $\vert \neutralSemicondState \rangle$ is the semiconductor's Dirac-sea ground state.
The full BSE for the exciton spectra at $T = 0$ is given by the following generalized eigenvalue problem \cite{hirata1999time,onida2002electronic,gruning2009exciton,sander2015beyond,maggio2016correlation,sander2017macroscopic}, for which we provide a path-integral derivation in App.\ \ref{app:derivationGenEVProblem}:
\begin{equation}
\label{eq:genEVProblem}
	\begin{bmatrix}
		H^{(\text{res})} & \bar{V} \\ \bar{V}^{*} & [H^{(\text{res})}]^{*} 
	\end{bmatrix} \! \begin{bmatrix}
		\Phi \\ \bar{\Phi}
	\end{bmatrix} = \omega \begin{bmatrix}
		\mathbb{I} & 0 \\ 0 & {-}\mathbb{I} 
	\end{bmatrix} \! \begin{bmatrix}
		\Phi \\ \bar{\Phi}
	\end{bmatrix} ,
\end{equation}
with an implied summation over internal bands and momenta.
The resonant part is $H^{\text{(res)}}_{cvc'v'}(\vmomentum, \vmomentum') = \delta_{\vmomentum \vmomentum'} \delta_{cc'} \delta_{vv'} (\epsilon^{c}_{\vmomentum} - \epsilon^{v}_{\vmomentum}) - \frac{1}{\mathcal{A}} V^{\mathrm{D}}_{cvc'v'}(\vmomentum, \vmomentum')$, while $\bar{V}_{cvc'v'}(\vmomentum, \vmomentum') = {-} \frac{1}{\mathcal{A}} V^{\mathrm{D}}_{cvv'c'}(\vmomentum, \vmomentum')$ is the resonant-antiresonant coupling.
Here, $\epsilon^{\alpha}_{\vmomentum}$ and $\vert u^{\alpha}_{\vmomentum} \rangle$ are the single-particle energies and eigenstates, and the so-called direct interaction reads \cite{wu2015exciton,maisel2023single}
\begin{equation}
\label{eq:directInteraction}
    V^{\mathrm{D}}_{\alpha \beta \alpha' \beta'}(\vmomentum, \vmomentum') = V(\vmomentum - \vmomentum') \langle u^{\alpha}_{\vmomentum} \vert u^{\alpha'}_{\vmomentum'} \rangle \langle u^{\beta'}_{\vmomentum'} \vert u^{\beta}_{\vmomentum} \rangle .
\end{equation}
For the model of Eq.~\eqref{eq:intro:singleParticleH}, $\alpha, \beta, \dots$ take values in $\{c,v\} \times \{{\uparrow},{\downarrow}\}$.
We see that $\bar{V}$ represents the ``vacuum-to-pair'' processes in the first row of Fig.\ \hyperref[fig:figIntro]{1b}, which break the individual $\mathrm{U}(1)$ symmetries.
In general one should also include the exchange interaction, but this vanishes here as we only consider excitons with zero total momentum and an electrostatic interaction that does not depend on orbital character \cite{qiu2015nonanalyticity,wu2015exciton,maisel2023single}.

The exciton states are classified in the orthogonal spin configurations $\vert c{\uparrow}, v{\uparrow} \rangle$ and $\vert c{\uparrow}, v{\downarrow} \rangle$, corresponding to antiparallel and parallel spins for the electron and the hole, respectively.
We have calculated the exciton energies of each subspace for values of $\varepsilon_{s}$ up to 7 and found that the ground-state energies are all larger than the gap, indicating a potential excitonic instability.
We define $\Delta E = E_{\uparrow \uparrow} - E_{\uparrow \downarrow}$ as the energy difference between the antiparallel and parallel configurations.
In the TDA, the antiparallel sector lies slightly higher in energy, with $\Delta E^{\text{TDA}}_{\ce{AsO}} \approx \SI{0.9 \pm 0.3}{\milli\electronvolt}$ and $\Delta E^{\text{TDA}}_{\text{MTCO}} \approx \SI{4.1 \pm 0.2}{\milli\electronvolt}$.
Remarkably, this ordering inverts under the inclusion of the resonant-antiresonant coupling, giving $\Delta E^{\text{BSE}}_{\ce{AsO}} \approx \SI{-0.32 \pm 0.08}{\milli\electronvolt}$ and $\Delta E^{\text{BSE}}_{\text{MTCO}} \approx \SI{-0.35 \pm 0.09}{\milli\electronvolt}$ \footnote{We have solved the BSE (with and without the TDA) in polar coordinates by discretizing the momentum magnitude for different numbers of points $N$.
The error intervals in the reported values of $\Delta E$ correspond to the uncertainties of the intercepts when extrapolating the obtained energies to the continuum by considering $E$ as a function of $1/N$.
We have used enough values of $N$ to accurately determine the sign of $\Delta E$.}.
This gives a first indication that condensation with antiparallel spins may be favored over the pure triplet channel found in Ref.~\cite{yang2024spin} in the presense of $\mathrm{U}(1)$-symmetric channels only, which we now verify by explicitly solving the mean-field equations.

\section{Mean-field theory}

\noindent
We now study the EI phase in these materials via mean-field theory at $T \ge 0$.
After a Hartree-Fock treatment of the interaction we identify the gap parameters
\begin{equation}
\label{eq:meanField:gapParams}
	\Delta^{\alpha \beta}_{\vmomentum} = {-} \frac{1}{\mathcal{A}} \sum_{\alpha' \beta' \vmomentum'} V^{\mathrm{D}}_{\alpha \beta \alpha' \beta'}(\vmomentum, \vmomentum') \rho^{\alpha' \beta'}_{\vmomentum'} ,
\end{equation}
with $\rho^{\alpha \beta}_{\vmomentum} = \langle \hat{\psi}^{\dagger}_{\beta \vmomentum} \hat{\psi}_{\alpha \vmomentum} \rangle - \delta_{\alpha \beta} \delta_{\alpha, v}$ the density matrix relative to the Dirac sea.
We find mean-field configurations by iterating over pairs $(\Delta, \rho)$ until the free energy has converged, using that $\langle \hat{\psi}^{\dagger}_{\beta \vmomentum} \hat{\psi}_{\alpha \vmomentum} \rangle = \sum_{\gamma} [\mathcal{U}^{\Delta}_{\vmomentum}]_{\alpha \gamma} N_{\mathrm{F}}(\omega^{\gamma}_{\vmomentum} - \mu) [\mathcal{U}^{\Delta}_{\vmomentum}]^{\dagger}_{\gamma \beta}$.
Here, $N_{\mathrm{F}}(x) = (\ec^{\beta x} + 1)^{-1}$ is the Fermi-Dirac distribution, $\mathcal{U}^{\Delta}_{\vmomentum}$ the matrix diagonalizing the mean-field Hamiltonian $\mathcal{H}^{\Delta}_{\alpha \beta}(\vmomentum) = \delta_{\alpha \beta} \epsilon^{\alpha}_{\vmomentum} + \Delta^{\alpha \beta}_{\vmomentum}$ with eigenvalues $\omega^{\alpha}_{\vmomentum}$, and $\mu$ the chemical potential which is found by imposing charge neutrality.

\begin{figure}[!t]
    \centering
    \includegraphics[width=\linewidth]{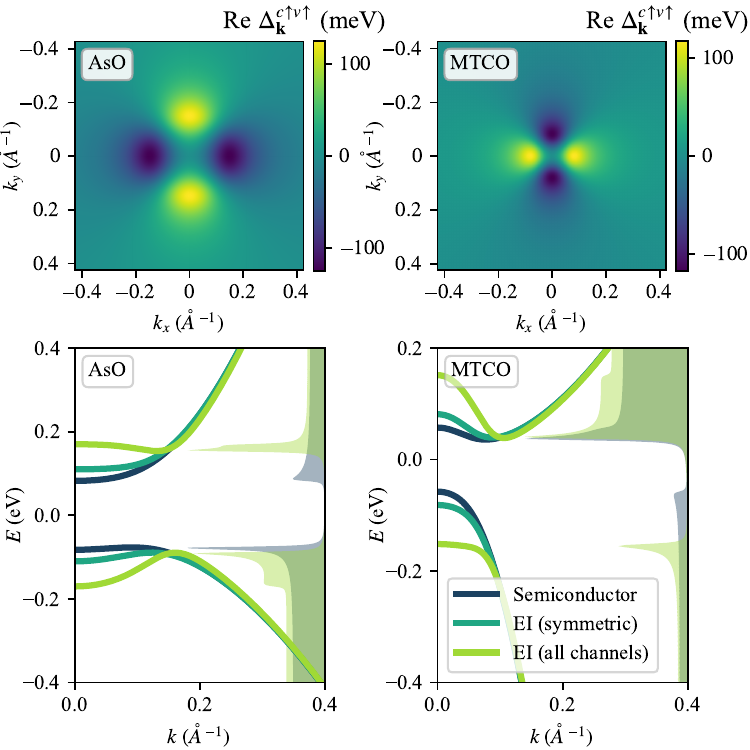}
    \caption{Real part of the gap function $\Delta^{c {\uparrow} v {\uparrow}}_{\vmomentum} \propto k_{-}^{2}$ (top) and quasiparticle dispersions (bottom) for both materials.
    We compare the dispersions in the presence of all channels and of $\mathrm{U}(1)$-symmetric processes only.
    In the former case the gap is topological and is widened by the diagonal gap parameters $\Delta^{c {\uparrow} c {\uparrow}}_{\vmomentum}$ and $\Delta^{v {\uparrow} v {\uparrow}}_{\vmomentum}$.
    The right side of the plots show the density-of-states profiles of the TI and of our $d + \iu d$ EI in arbitrary units.
    Here we have used $\varepsilon_{s} = 4$ for \ce{AsO} and $\varepsilon_{s} = 7$ for MTCO.}
    \label{fig:meanFieldResults}
\end{figure}

For large $\varepsilon_{s}$ we recover the underlying TI at all temperatures, while for small enough $\varepsilon_{s}$ we find an EI.
The transition as a function of $\varepsilon_{s}$ is of first order, compatible with the fact that the two phases share the same symmetries due to the presence of the $\mathrm{U}(1)$-breaking terms.
The EI obtained here is more robust than in the number-conserving approximation, with critical dielectric constants $\varepsilon^{\mathrm{AsO}}_{s, c} \approx 7.1$ and $\varepsilon^{\mathrm{MTCO}}_{s, c} \approx 11.6$ at $T = 0$, which are higher than those reported in Ref.~\cite{yang2024spin}.
Furthermore, they exist in a metastable state up to $\varepsilon^{\text{AsO}}_{s,m} \approx 9.2$ and $\varepsilon^{\text{MTCO}}_{s, m} \approx 13.7$.
In Fig.~\ref{fig:meanFieldResults} we present the excitonic gap parameter and quasiparticle dispersions at $T = 0$.
We obtain $\Delta^{\alpha \beta}_{\vmomentum} = 0$ for all channels where the spins of bands $\alpha$ and $\beta$ are antiparallel.
By contrast, $\Delta^{c \uparrow v \uparrow}_{\vmomentum} = \Delta(\mmomentum) k_{-}^{2}$ with $\Delta(\mmomentum)$ a decreasing function of $k$.
Meanwhile, the diagonal components are of $s$-wave type and satisfy $\Delta^{c \uparrow c \uparrow}_{\vmomentum} = {-} \Delta^{v \uparrow v \uparrow}_{\vmomentum}$.
The gap parameters in the spin-down subspace are the complex conjugates of the spin-up ones and the density matrix has the same form.

These results can be contrasted with Ref.~\cite{yang2024spin}, where the gap parameter is an $s$-wave triplet in the channels $(c{\uparrow}, v{\downarrow})$ and $(c{\downarrow}, v{\uparrow})$, with $\Delta_{\vmomentum}^{c \uparrow v \uparrow} = \Delta_{\vmomentum}^{c \downarrow v \downarrow} = 0$.
The discrepancy arises due to the $\mathrm{U}(1)$-symmetric approximation for the interaction matrix where only $V^{\mathrm{D}}_{cvc'v'}$ is kept, which is not permissible here owing to the phase winding of $H_{0}(\vmomentum)$.
By dressing the bare electrostatic potential with momentum-dependent form factors, this phase winding imbues the bare interaction with a channel-dependent angular structure.
We illustrate this by highlighting the angular dependence of the matrix elements entering the generalized eigenvalue problem of Eq.\ \eqref{eq:genEVProblem}:
\begin{subequations}
	\begin{align}
		\label{eq:VDU1symm}
		V_{c \uparrow v \uparrow c \uparrow v \uparrow}^{\mathrm{D}}(\vmomentum, \vmomentum') &= v(k, k', \phi_{\vmomentum} - \phi_{\vmomentum'}) , \\
		\label{eq:VDU1br}
		V_{c \uparrow v \uparrow v \uparrow c \uparrow}^{\mathrm{D}}(\vmomentum, \vmomentum') &= \bar{v}(k, k', \phi_{\vmomentum} - \phi_{\vmomentum'}) \hspace{0.25mm} \ec^{{-} 4 \iu \phi_{\vmomentum'}} ,
	\end{align}	
\end{subequations}
where $\phi_{\vmomentumtransfer}$ is the polar angle of $\vmomentumtransfer$.
The form of the functions $v$ and $\bar{v}$ is irrelevant for this discussion, the only important point is that they only depend on the angle difference.
If only the $\mathrm{U}(1)$-symmetric \eqref{eq:VDU1symm} is included in Eq.~\eqref{eq:meanField:gapParams}, then there is a self-consistent solution for $\rho^{c\uparrow v\uparrow}_{\vmomentum}$ proportional to $\ec^{\iu m \phi_{\vmomentum}}$ for any integer $m$, as can be seen by shifting $\phi_{\vmomentum'}$ in the angular integral.
Under this approximation, the solution with $m = 0$ has the lowest energy.
However, once the $\mathrm{U}(1)$-breaking coupling of Eq.\ \eqref{eq:VDU1br} is included as well, the angular integral with $\rho^{c \uparrow v \uparrow}_{\vmomentum'} \propto \ec^{\iu m \phi_{\vmomentum'}}$ leads to a new term proportional to $\ec^{{-}\iu (m + 4) \phi_{\vmomentum}}$ due to the lone factor $\ec^{{-} 4 \iu \phi_{\vmomentum'}}$ accompanying $\bar{v}$ \footnote{We note that the $\mathrm{U}(1)$-conserving coupling of Eq.\ \eqref{eq:VDU1symm} is integrated against $\rho^{c \uparrow v \uparrow}_{\vmomentum'} \propto \ec^{\iu m \phi_{\vmomentum'}}$, while the $\mathrm{U}(1)$-breaking coupling of Eq.\ \eqref{eq:VDU1br} is integrated against $\rho^{v \uparrow c \uparrow}_{\vmomentum'} = (\rho^{c \uparrow v \uparrow}_{\vmomentum'})^{*} \propto \ec^{{-}\iu m \phi_{\vmomentum}}$, as follows from the order of the indices in Eq.\ \eqref{eq:meanField:gapParams}}.
Consequently, a pure angular momentum $m = 0$ does not give rise to a self-consistent solution, but $m = {-} 2$ does.
This continues to be the case after a similar analysis is performed for all channels present in Eq.\ \eqref{eq:meanField:gapParams}.
Another important point is that processes such as those of the second row of Fig.\ \hyperref[fig:figIntro]{1b} couple the diagonal and off-diagonal elements of the density matrix.
In other words, a nonzero $\rho^{c{\uparrow}c{\uparrow}}_{\vmomentum}$ always leads to a nonzero $\Delta^{c{\uparrow}v{\uparrow}}_{\vmomentum}$.
Similar considerations hold for the spin-down subspace.
The result is that $\Delta_{\vmomentum}^{c \uparrow v \uparrow} = \Delta_{\vmomentum}^{c \downarrow v \downarrow} = 0$ is not possible in the EI in the presence of all channels while $\Delta_{\vmomentum}^{c \uparrow v \downarrow} = \Delta_{\vmomentum}^{c \downarrow v \uparrow} = 0$ is permissible, meaning that the spin-$\mathrm{U}(1)$ and time-reversal symmetries are both preserved.
Hence, including all available interaction channels into the self-consistent mean-field calculation can dramatically influence the ground state.
Our results are robust as long as $V(\vmomentumtransfer)$ depends solely on $|\vmomentumtransfer|$ and the potential in coordinate space has a long-range tail.
It is important to realize that a contact potential can lead to artificious results in the presence of nontrivial winding of the single-particle states, as the latter will in general couple to the nonzero angular-momentum components of $V(\vmomentumtransfer)$.
Despite the fact that we work in the long-wavelength limit, the last statement also translates to lattice models: in the presence of topology it is in general not enough to consider the on-site repulsion only, and one should at least take nearest-neighbor interactions into account.

The specific pairing symmetry and spin structure of the gap are model-dependent.
Here we obtain a chiral $d$-wave state due to the hybridizing terms $\mmomentum_{\pm}^{2}$ in the single-particle Hamiltonian of Eq.\ \eqref{eq:intro:singleParticleH}, which propagate into the interaction matrix.
If this were replaced with a winding $w$, it would lead to interband gap parameters proportional to $(\mmomentum_{x} \pm \iu \mmomentum_{y})^{w}$.
A $p$-wave EI arises for $w = 1$, which corresponds to a massive Dirac model.
Note that a $p$-wave singlet EI is allowed because electrons and holes are nonidentical particles with less restrictive pairing symmetries than ordinary superconductors.
In this sense, the EI is akin to a multiband BCS state.

\section{Topological properties and experiments}

\noindent
To study the topology of this EI we firstly focus on the Chern numbers of the quasiparticle bands, which remain well-defined due to the block-diagonal form of the mean-field Hamiltonian in the EI phase.
The quasiparticle dispersions read $\omega^{\pm}_{\vmomentum} = \frac{1}{2}\big[\varepsilon^{c}_{\vmomentum} + \varepsilon^{v}_{\vmomentum} \pm \sqrt{\smash[b]{(\varepsilon^{c}_{\vmomentum} - \varepsilon^{v}_{\vmomentum})^{2} + 4 \vert \Delta^{cv}_{\vmomentum} \vert^{2}}}\big]$, where $\varepsilon^{\alpha}_{\vmomentum} = \epsilon^{\alpha}_{\vmomentum} + \Delta^{\alpha \alpha}_{\vmomentum}$.
Here we implicitly take equal spins for $c$ and $v$ and omit the label.
Even in the presence of nontrivial gap parameters, the Chern numbers can be analytcially computed by integrating the Berry curvature in our continuum model, and the result is
\begin{equation}
	\mathcal{C}^{\pm}_{\sigma}[\Delta] = \pm \sigma \operatorname{sgn}(\varepsilon^{c}_{\vec{0}} - \varepsilon^{v}_{\vec{0}}) \operatorname{sgn} \lambda .
\end{equation}
The parameter $\lambda$ was introduced in Eq.~\eqref{eq:intro:singleParticleH} and in our case is positive for both systems.
The Chern numbers become opposite to the TI ones whenever the quasiparticle gap is inverted, i.e., $\Delta^{cc}_{\vec{0}} = {-} \Delta^{vv}_{\vec{0}} < {-}(\epsilon^{c}_{\vec{0}} - \mu) < 0$.
As shown in Fig.\ \hyperref[fig:figIntro]{1d}, this is the case in a large region of the parameter space $(\varepsilon_{s}, T)$, and thus we can distinguish the EI and the TI via different topological invariants even though they share the same symmetries.

To further understand the topological properties we connect the density matrix with the BCS-like ground state $\vert \Psi_{\mathrm{EI}} \rangle = \prod_{\vmomentum \sigma} (u^{\sigma}_{\vmomentum} + v^{\sigma}_{\vmomentum} \hat{\psi}^{\dagger}_{c \vmomentum \sigma} \hat{\psi}_{v \vmomentum \sigma}) \vert \neutralSemicondState \rangle$ via $\rho^{c \sigma , c \sigma}_{\vmomentum} = {-} \rho^{v \sigma , v \sigma}_{\vmomentum} = \vert v^{\sigma}_{\vmomentum} \vert^{2}$ and $\rho^{c \sigma , v \sigma}_{\vmomentum} = v^{\sigma}_{\vmomentum} (u^{\sigma}_{\vmomentum})^{*}$.
Defining $g^{\sigma}_{\vmomentum} = v^{\sigma}_{\vmomentum} / u^{\sigma}_{\vmomentum}$ we find $g^{\sigma}_{\vmomentum} \propto (k_{x} + \iu \sigma k_{y})^{-2}$ around the origin.
This corresponds to the weak-pairing abelian phase described in Refs.~\cite{senthil1999spin,read2000paired}, and following similar arguments we conclude that there must exist fermionic bound states at the edges of a sample, which will be helical due to the preserved time-reversal symmetry.
However, the same is true for the TI, so experimentally it would be difficult to distinguish between these two phases on the basis of the density of states at the edge of a sample on a homogeneous substrate.
Nevertheless, we can exploit the opposite winding numbers of the high- and low-$\varepsilon_{s}$ phases to propose and experimental setup capable of distinguishing between the TI phase and the $d + \iu d$ EI as follows.

\begin{figure}[!t]
    \centering
    \includegraphics[width=\linewidth]{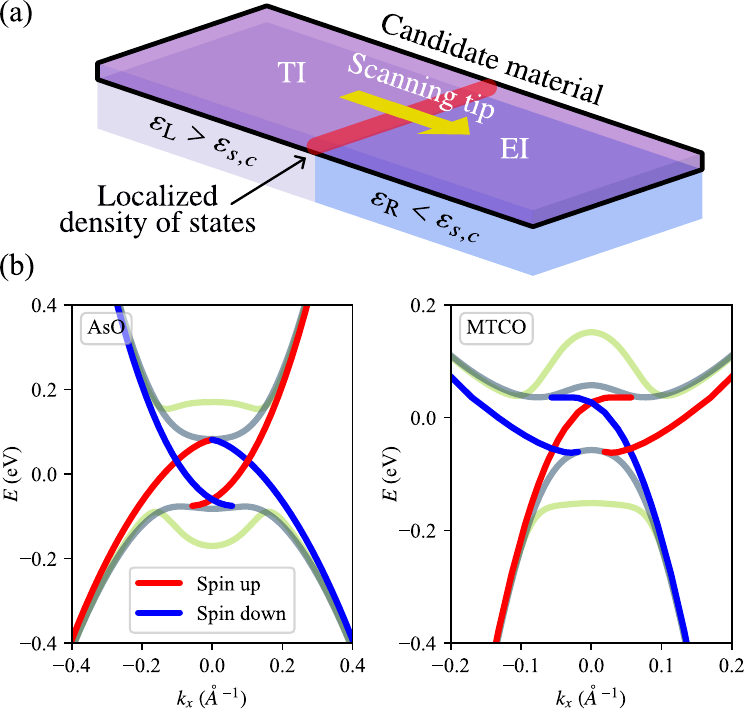}
    \caption{\textbf{(a)} Experimental setup described in the main text.
    The left side of the candidate material lies on a substrate of high dielectric constant, while the substrate on the right side has low $\varepsilon_{s}$.
    We then have different topological phases at both sides, in particular $x \gg 0$ contains the EI phase.
    This leads to localized topological states at the interface between both regions which can be detected experimentally by scanning the density of states across the interface at energies lying inside the band gap.
    \textbf{(b)} Approximate band structure of the proposed system showing two pairs of helical edge modes crossing the band gap.
    While not shown here, their wave functions are localized at the interface and decay exponentially away from it.}
    \label{fig:interfaceStates}
\end{figure}

We assume that the candidate material is placed on top of a substrate with a position-dependent dielectric constant, which we take as $\varepsilon(y < 0) = \varepsilon_{\mathrm{L}} > \varepsilon_{s,c}$ and $\varepsilon(y > 0) = \varepsilon_{\mathrm{R}} < \varepsilon_{s,c}$.
For $y \gg 0$ the system is in the EI phase, while for $y \ll 0$ it lies in the TI regime.
The winding numbers of these phases differ by two, and thus by the bulk-boundary correspondence we expect two pairs of helical edge states located at the interface between the two regions.
These can be found by matching the solutions of $H_{0}$ for $y < 0$ with those of $\mathcal{H}^{\Delta} = H_{0} + \Delta$ for $y > 0$, where $\Delta$ is a matrix containing the nontrivial gap parameters.
While solving the full model is difficult due to the complicated $\vmomentum$-dependence of the many-body gap, to obtain a qualitative picture we can approximate the Hamiltonian for $y > 0$ as $H_{0}$ with the substitution $\lambda \rightarrow \lambda + \Delta^{cc}_{\vec{0}}$.
This preserves the crucial effect of the nontrivial gap parameters in $\Delta$, which is to invert the gap in the topological EI regime, while preserving the small- and large-$\vmomentum$ behaviors of the quasiparticle dispersions and wave functions.
Meanwhile, the behavior at intermediate momenta is irrelevant for our purposes because it may be obtained via a smooth deformation.
We then change $k_{y} \rightarrow {-} \iu \partial_{y}$ and perform a Jackiw-Rebbi-like calculation \cite{jackiw1976solitons} yielding the bands shown in Fig.\ \hyperref[fig:interfaceStates]{3b}, which conform to our expectations.
We note that the same can be achieved by placing the sample on a homogeneous substrate with an additional dielectric on top of half the sample in such a way that the total effective dielectric constants at both sides still satisfy our requirements, which may be more feasible experimentally.

\section{BKT physics}

\noindent
Being a condensate state, the EI carries a superfluid weight $D_{s}$, defined as the coefficient of the phase-gradient term in the associated Ginzburg-Landau action.
In fact, due to the presence of the $\mathrm{U}(1)$-breaking terms, the superfluid weight carried by the excitonic condensate can never fully vanish at nonzero temperatures.
Hence, the TI must always have an extremely small, but technically nonzero superfluid fraction.
It is important to emphasize that the presence of a nonzero $D_{s}$ does not automatically imply the presence of superfluid \emph{flow}.
Indeed, the presence or absence of the latter will ultimately depend on the relative strength between the $\mathrm{U}(1)$-conserving and $\mathrm{U}(1)$-breaking processes.
When the latter are much smaller than the former, superfluidity may take place as a transient phenomenon \cite{nagaoka1975phase}.
In view of the significant influence of the $\mathrm{U}(1)$-breaking terms on the pairing structure, it is possible that this is not the case here.
However, a full analysis of the superfluid properties of this unconventional EI phase lies outside the scope of this work.
We do note that, given that we are considering monolayer systems, the superfluid response should be experimentally probed in terms of a heat current instead of an electric current.

\begin{figure}[!t]
    \centering
    \includegraphics[width=\linewidth]{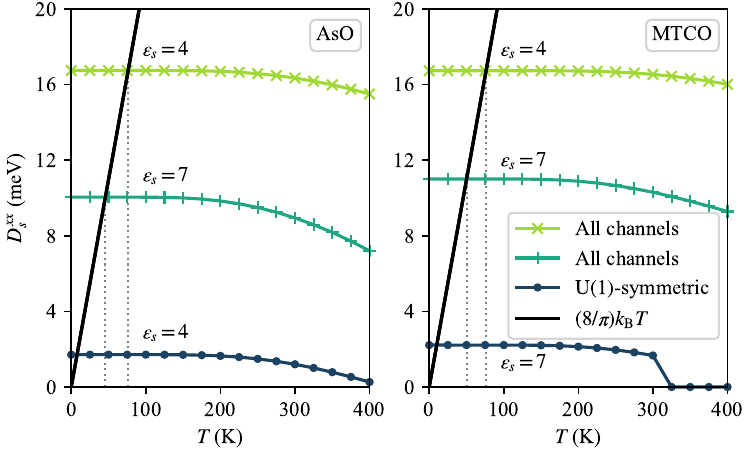}
    \caption{Superfluid weight as a function of temperature and visualization of $T_{\mathrm{BKT}}$ via the Nelson-Kosterlitz criterion for substrate dielectric constants $\varepsilon_{s} = 4, 7$.
    The latter is slightly higher than that of the hexagonal boron nitride and gives a $T_{\mathrm{BKT}}$ of around $\SI{50}{\kelvin}$.
    Thus, we identify these systems as candidates for hosting BKT physics at relatively high temperatures on a realistic substrate.
    We also show $D_{s}$ for the EI with only $\mathrm{U}(1)$-symmetric channels included, whose $T_{\mathrm{BKT}}$ is of at most a few Kelvin.}
    \label{fig:DsvsT}
\end{figure}

We have computed the bare superfluid weight \cite{peotta2015superfluidity,liang2017band,kitamura2022quantum,kitamura2022superconductivity} for our $d$-wave EI as well as for the $s$-wave ground state in the $\mathrm{U}(1)$-conserving approximation.
The results are plotted in Fig.\ \ref{fig:DsvsT} and show that $D_{s}$ for our topological $d$-wave EI is dramatically enhanced with respect to the $\mathrm{U}(1)$-symmetric case.
While superfluid flow may be absent in the EI due to the symmetry-breaking interactions, there will be a BKT-like crossover from a state of bound vortex-antivortex pairs to one of (almost) free vortices.
To estimate its temperature $T_{\mathrm{BKT}}$ we intersect the bare superfluid weight $D_{s}$ with the well-known universal line.
The use of the unrenormalized $D_{s}$ for estimating $T_{\mathrm{BKT}}$ is known as the Nelson-Kosterlitz criterion and should provide a rather reliable upper bound to the crossover temperature \cite{nelson1977universal,furutani2024amplitude}.
The phase boundary obtained with this approach as a function of the dielectric constant spans temperatures in the range of $\SI{25}{\kelvin}$ to $\SI{100}{\kelvin}$ for both materials.
This is very high compared to those of superfluid helium or conventional superconductors.
For this calculation we have ignored the self-consistency of the gap parameters derived from the microscopic model and instead treated them as fixed, as is often done for simplicity.
Also, we only consider the conventional part of the superfluid weight, which is warranted because the dispersions are not flat \cite{tam2024geometry}.
We further note that the traditional BKT theory can in principle be employed in $\mathrm{U}(1)$-invariant systems only.
Our calculation thus assumes that the strength of the $\mathrm{U}(1)$-breaking terms is relatively small compared to the number-conserving channels, which we expect to be the case since they all involve overlaps between different bands.
From this perspective, the effect of adding the $\mathrm{U}(1)$-breaking channels is to make the $d + \iu d$ EI the only possible solution, and its superfluid weight is what we compute here.

\section{Discussion and outlook}

\noindent
In summary, we have studied the EI phase in two recently proposed candidate materials in the presence of symmetry-breaking interaction channels which arise due to the inherent nonlocalizability of the Wannier functions associated to topological bands.
While in the TI phase these channels merely refine the calculation of exciton spectra, in an EI phase they can be paramount to obtain the correct ground-state configuration.
Due to the $\mathrm{U}(1)$-breaking terms we obtain an unconventional exciton pairing whose angular momentum is determined by the phase winding of the free electrons and in the cases considered here corresponds to $d + \iu d$ order parameters.
The corresponding EI is topologically nontrivial and distinct from the TI phase, a property that we have employed to propose an experimental setup capable of distinguishing between the two via density-of-states measurements in the considered systems.
The EI found here also displays a large superfluid weight leading to a high-temperature BKT crossover in the vicinity of $\SI{75}{\kelvin}$ in both materials on experimentally achievable substrates.

The results reported in this work all refer to condensation with vanishing total exciton momentum, $\vMomentum = \vec{0}$.
However, we expect the $\mathrm{U}(1)$-breaking terms to also play a role in the condensation at nonzero $\vMomentum$, where previous research has shown that even in the semiconductor state it can be important to go beyond them \cite{sander2015beyond}.
A first result in this regard is that the Larkin-Ovchinnikov phase $\Delta^{cv}_{\vMomentum_{0}}(\vec{x}) \propto \cos (\vMomentum_{0} \vec{\cdot} \vec{R})$ (with $\vec{R}$ the coordinate-space center of mass) should appear instead of the Fulde-Ferrel phase $\Delta^{cv}_{\vMomentum_{0}}(\vec{R}) \propto \ec^{\iu \vMomentum_{0} \vec{\cdot} \vec{R}}$, as the $\mathrm{U}(1)$-breaking channels couple the $\pm \vMomentum$ components of the density matrix.
Nonzero-$\vMomentum$ exciton condensation should naturally occur in systems where the single-particle band structure has a pronounced camelback shape, such as monolayer $1T'$-\ce{WSe2} \cite{dong2025topological}.
This material would be an ideal candidate to explore the effects associated with these symmetry-breaking channels for $\vMomentum \neq \vec{0}$ by solving the excitonic gap equation.

The dressing of the effective electron-electron interaction with hybridization effects leading to $\mathrm{U}(1)$-breaking channels is a universal mechanism.
Their influence on the pairing of fermions will be important whenever the effects of Berry curvature are significant, or equivalently, when there is substantial overlap between the wave functions of electrons in different bands.
This is especially true when some of the involved bands have a nonzero Chern number.
As a concrete example, in Ref.\ \cite{maisel2023single} we studied excitons in a Bernevig-Hughes-Zhang model which can be tuned from a trivial state to a topological one.
Our results show that Berry-curvature effects are crucial in the topological regime, but that a simple effective-mass description which disregards them would suffice in the trivial case.
More generally, we note that the Berry curvature can also play a role even when there is no overall band topology.
An example of this are transition-metal dichalcogenides, where the Berry curvatures at the $K$ and $K'$ points cancel each other, but are still significant within each individual valley.

Finally, while our work focuses on the EI state, the considerations above indicate that similar physics can be relevant for other kinds of fermion pairing.
In particular, it would be of interest to study how the $\mathrm{U}(1)$-breaking interactions affect the pairing of electrons in multiband superconductors, where breaking of the \emph{total} charge conservation can still occur.
In this regard we emphasize that our model calculation of the EI properties naturally leads to high-angular-momentum excitonic pairing due to the winding of the single-particle Hamiltonian.
It is reasonable to assume that a similar effect could take place for Cooper pairs, potentially leading to $p$- or $d$-wave superconductivity.
More generally, our findings suggest that unconventional pairing should be rather ubiquitous when topological bands are involved and may go hand in hand with a large superfluid weight.
We hope this will further motivate the search of exotic phases of matter in systems hosting both topological properties and long-range interactions.

\section{Acknowledgments}

\noindent
We are indebted to the authors of Ref.~\cite{yang2024spin} for providing the numerical values of the $\vmomentum \vec{\cdot} \vec{p}$ model parameters for \ce{AsO} and MTCO.
We also thank Daniel Vanmaekelbergh and Ingmar Swart for useful discussions, as well as Lumen Eek for a suggestion regarding the experimental proposal.
This work is supported by the Delta-ITP consortium and by the research program \textit{QuMat--Materials for the Quantum Age}.
These are programs of the Netherlands Organisation for Scientific Research (NWO) and the Gravitation pogram, respectively, which are funded by the Dutch Ministry of Education, Culture, and Science (OCW).

\appendix

\section{Coulomb interaction in a $\vmomentum \vec{\cdot} \vec{p}$ model}
\label{app:CoulombIntkDotpModel}

\noindent
Here we show that in a $\vmomentum \vec{\cdot} \vec{p}$ model, the Coulomb interaction only conserves the electronic degrees of freedom at the interaction vertex when it is written in the basis of states at the momentum around which the expansion is performed.
While this may seem obvious, many studies in the literature first diagonalize the free electron Hamiltonian and then add a single interband interaction term between the conduction and valence bands.
This neglects important overlap between Bloch factors which are especially crucial when the bands have a nonzero Chern number.

Let us start from the following Hamiltonian for interacting electrons in a solid:
\begin{equation}
\label{eq:chMiscConc:generalInteractingHamiltonianMicroscopic}
	\begin{split}
		&\hat{H} = \sum_{\sigma \sigma'} \int_{\vec{x}} \hat{\psi}^{\dagger}_{\sigma}(\vec{x}) H^{0}_{\sigma \sigma'}(\vec{x}, {-} \iu \vec{\nabla}) \hat{\psi}_{\sigma'}(\vec{x}) \\
		&\! + \frac{1}{2} \sum_{\sigma \sigma'} \int_{\vec{x} \vec{x}'} \hat{\psi}^{\dagger}_{\sigma}(\vec{x}) \hat{\psi}^{\dagger}_{\sigma'}(\vec{x}') V(\vec{x} - \vec{x}') \hat{\psi}_{\sigma'}(\vec{x}') \hat{\psi}_{\sigma}(\vec{x}) ,
	\end{split}
\end{equation}
where $\int_{\vec{x}} = \int \mathrm{d}^{2} x$.
Here, $H_{0}$ is the microscopic Hamiltonian containing the usual quadratic momentum term and the periodic lattice potential satisfying $H_{0}(\vec{x} + \vec{R}, {-} \iu \vec{\nabla}) = H_{0}(\vec{x}, {-} \iu \vec{\nabla})$ for any lattice vector $\vec{R}$.
By including a spin index we also allow for effects such as spin-orbit coupling.
The field operators have a spin and position label, as these are the fundamental degrees of freedom of the electrons.
The Hamiltonian above does not assume anything about the system and must be valid as a starting point for any calculation in crystalline solids.
For now we forget about the interaction term and focus on the quadratic part.
From Bloch's theorem, we know that we can write the eigenstates of this Hamiltonian containing a periodic potential as
\begin{equation}
  \psi_{\alpha \vmomentum}(\vec{x}) = \ec^{\iu \vmomentum \vec{\cdot} \vec{x}} u_{\alpha \vmomentum}(\vec{x}) ,
\end{equation}
where $\vmomentum$ is the crystal momentum lying in the first Brillouin zone and $u_{\alpha \vmomentum}(\vec{x})$ has the periodicity of the lattice.
The Bloch functions $\psi_{\alpha \vmomentum}(\vec{x})$ are normalized over the entire crystal, while the Bloch factors $u_{\alpha \vmomentum}(\vec{x})$ satisfy a normalization over the unit cell.
When $H$ mixes spins, the Bloch wave functions become two-component spinors; henceforth we omit the spin index everywhere by treating everything as matrices or vectors as appropriate.

To make contact with the $\vmomentum \vec{\cdot} \vec{p}$ formalism we now wish to derive an effective model that describes the system in the vicinity of some wave vector $\vmomentum_{0}$ of choice.
Here we choose $\vmomentum_{0} = \vec{0}$, but everything can be straightforwardly generalized to a nonzero $\vmomentum_{0}$.
Using the above Bloch wave functions as an {\it Ansatz} leads to the following equation for the Bloch factors:
\begin{equation}
  H_{0}(\vec{x}, {-}\iu \vec{\nabla} + \vmomentum) u_{\alpha \vmomentum}(\vec{x}) = E_{\alpha \vmomentum} u_{\alpha \vmomentum}(\vec{x}) .
\end{equation}
Let us assume that we have solved this equation at the $\Gamma$ point and denote the solutions by $\phi_{a}(\vec{x})$.
The index $a$, as opposed to $\alpha$, stands for the eigenstates at the $\Gamma$ point which will serve as the basis for the $\vmomentum \vec{\cdot} \vec{p}$ model to be introduced shortly.
The functions $\phi_{a}$ form a complete set for the functions with the periodicity of the lattice.
Next, we use them to expand the field operators as
\begin{equation}
\label{eq:chMiscConc:expansionFieldOperatorsGammaPoint}
  \hat{\psi}(\vec{x}) \propto \sum_{a \vmomentum} \ec^{\iu \vmomentum \vec{\cdot} \vec{x}} \phi_{a}(\vec{x}) \hat{\psi}_{a \vmomentum}
\end{equation}
up to an irrelevant normalization.
Here, $\hat{\psi}_{a \vmomentum}$ annihilates an electron in state $a$ at wave vector $\vmomentum$.
Note that the expansion is not in terms of eigenstates of $H_{0}$, but is valid nonetheless because the functions $\ec^{\iu \vmomentum \vec{\cdot} \vec{x}} \phi_{a}(\vec{x})$ still form a complete orthonormal set.
Indeed, the inverse relation is \footnote{To show this one decomposes the real-space integral as $\int \mathrm{d}^{2} x \, f(\vec{x}) = \sum_{\vec{R}} \int_{\mathrm{UC}} \mathrm{d}^{2} x \, f(\vec{x} + \vec{R})$, where the sum runs over all lattice vectors, and then uses $\sum_{\vec{R}} \ec^{\iu (\vmomentum - \vmomentum') \vec{\cdot} \vec{R}} \propto \delta_{\vmomentum \vmomentum'}$ as well as the periodicity and the unit-cell normalization of the $\phi_{a}$'s.}
\begin{equation}
  \hat{\psi}_{a \vmomentum} \propto \int_{\vec{x}} \ec^{{-} \iu \vmomentum \vec{\cdot} \vec{x}} \phi_{a}^{\dagger}(\vec{x}) \hat{\psi}(\vec{x}) .
\end{equation}
Plugging this expansion for the field operators into the quadratic Hamiltonian yields
\begin{equation}
\label{eq:H0AfterExpansionHiikDef}
	\hat{H}_{0} = \sum_{aa'} \bigg[\int_{\vec{x}, \mathrm{UC}} \phi^{\dagger}_{a}(\vec{x}) H_{0}(\vec{x}, {-} \iu \vec{\nabla} + \vmomentum) \phi_{a'}(\vec{x})\bigg] \hat{\psi}^{\dagger}_{a \vmomentum} \hat{\psi}_{a' \vmomentum} ,
\end{equation}
where the integral is performed over the unit cell.
The term inside the square brackets is the original Hamiltonian now written in the basis of Bloch functions at the $\Gamma$ point and will be denoted by $H^{0}_{aa'}(\vmomentum)$.

Now let us consider the interaction term which we have previously neglected.
Plugging the expansion \eqref{eq:chMiscConc:expansionFieldOperatorsGammaPoint} into the last term of Eq.~\eqref{eq:chMiscConc:generalInteractingHamiltonianMicroscopic} leads to
\begin{align}
\label{eq:chMiscConc:interactionTermGammaExpansion}
	\hat{V} = \; &\sum_{\{a\}} \bigg[ \int_{\vec{x}, \mathrm{UC}} \phi^{\dagger}_{a_{1}}(\vec{x}) \phi_{a_{4}}(\vec{x}) \bigg] \bigg[ \int_{\vec{x}', \mathrm{UC}} \phi^{\dagger}_{a_{2}}(\vec{x}') \phi_{a_{3}}(\vec{x}') \bigg] \nonumber \\
	&\times \sum_{\mathclap{\vmomentum \vmomentum' \vec{q}}} V(\vec{q}) \hat{\psi}_{a_{1} \vmomentum+\vec{q}}^{\dagger} \hat{\psi}_{a_{2} \vmomentum'- \vec{q}}^{\dagger} \hat{\psi}_{a_{3} \vmomentum'} \hat{\psi}_{a_{4} \vmomentum} .
\end{align}
The terms in brackets simply give $\delta_{a_{1} a_{4}} \delta_{a_{2} a_{3}}$ due to the unit-cell normalization of the basis of wave functions at the $\Gamma$ point.
The original Hamiltonian has then been transformed into
\begin{equation}
\label{eq:chMiscConc:HamiltonianBasisGamma}
  \begin{split}
  	\hat{H} &= \sum_{aa'} \sum_{\vmomentum} \hat{\psi}^{\dagger}_{a \vmomentum} H^{0}_{aa'}(\vmomentum) \hat{\psi}_{a' \vmomentum} \\
  	&+ \frac{1}{2 \mathcal{A}} \sum_{aa'} \sum_{\mathclap{\vmomentum \vmomentum' \vec{q}}} V(\vec{q}) \hat{\psi}_{a \vmomentum+\vec{q}}^{\dagger} \hat{\psi}_{a' \vmomentum'- \vec{q}}^{\dagger} \hat{\psi}_{a' \vmomentum'} \hat{\psi}_{a \vmomentum} ,
  \end{split}
\end{equation}
with $H^{0}_{aa'}(\vmomentum)$ being the term in brackets in Eq.\ \eqref{eq:H0AfterExpansionHiikDef}.
We note that the spin dependence is in general encoded into $H_{aa'}(\vmomentum)$ via the two-component spinors $\phi_{a}(\vec{x})$.
If the original Hamiltonian does not mix spins, then we may straightforwardly restore the spin labels in the above.
Note that the form of the interaction term in Eq.~\eqref{eq:chMiscConc:HamiltonianBasisGamma} arises only because we employ $\vmomentum$-independent Bloch factors in the expansion of the field operators \eqref{eq:chMiscConc:expansionFieldOperatorsGammaPoint}.
Had we used the Bloch eigenbasis $\psi_{\alpha \vmomentum}(\vec{x})$ in the expansion, we would obtain a diagonal single-particle Hamiltonian $H^{0}_{\alpha \beta}(\vmomentum) = \delta_{\alpha \beta} E_{\alpha \vmomentum}$ at the expense of overlaps between Bloch factors of different bands and momenta in the integrals of Eq.~\eqref{eq:chMiscConc:interactionTermGammaExpansion}.
This also correct, but in this case we may not use the completeness relation of the Bloch factors because they involve different momenta.
The overlaps must then be kept in the interaction term, precisely leading to the form factors that otherwise arise when diagonalizing the single-particle part of $\hat{H}$ in Eq.\ \eqref{eq:chMiscConc:HamiltonianBasisGamma}.

We stress that so far this is exact and describes exactly the same system as the starting model as long as $\vmomentum$ is taken inside the first Brillouin zone and all basis functions $\phi_{a}(\vec{x})$ are kept.
In practice one is often interested in the physics around the Fermi surface.
In this case it is allowed to keep only the basis elements with $\Gamma$-point energies closest to this point and disregard the states that lie further away.
The wave vector $\vmomentum$ should then be close to zero, but in reality we may now allow for an infinite range of $\vmomentum$ by assuming that all physical quantities of interest will decay fast enough within the region of validity of the truncated model.
The single-particle Hamiltonian in the truncated basis and with an infinite range of $\vmomentum$ is then denoted by $H^{0,\mathrm{eff}}_{aa'}(\vmomentum)$ and is nothing but the $\vmomentum \vec{\cdot} \vec{p}$ Hamiltonian.
For $\vmomentum = \vec{0}$ it gives the known energy of the state $\phi_{a}(\vec{x})$ at the $\Gamma$ point.
Diagonalizing $H_{0,\mathrm{eff}}(\vmomentum)$ for $\vmomentum \neq \vec{0}$ gives an approximation to the bands close to $\Gamma$ and allows one to construct the Bloch wave functions in the vicinity of this point.

We can now construct an effective field theory close to the Fermi surface by writing
\begin{equation}
  \sum_{aa'} \hat{\psi}^{\dagger}_{a \vmomentum} H^{0,\mathrm{eff}}_{aa'}(\vmomentum) \hat{\psi}_{a' \vmomentum} = \int_{\vec{x}} \hat{\psi}^{\dagger}_{a}(\vec{x}) H^{0,\mathrm{eff}}_{aa'}({-} \iu \vec{\nabla}) \hat{\psi}_{a'}(\vec{x}) ,
\end{equation}
where the sum over $\Gamma$-point states is now understood to be restricted to the energy range of interest around the $\Gamma$ point.
Here we have introduced an effective set of field operators $\hat{\psi}_{a}(\vec{x})$ which contain the band index at the $\Gamma$ point.
The change from $\vmomentum$ to ${-} \iu \vec{\nabla}$ is allowed because now $\vmomentum$ is regarded as a continuous, unbounded variable.
Thus we end up with
\begin{equation}
\label{eq:finalFormHeffFieldTheory}
	\begin{split}
		&\hat{H}_{\mathrm{eff}} = \sum_{aa'} \int_{\vec{x}} \hat{\psi}^{\dagger}_{a}(\vec{x}) H^{0,\mathrm{eff}}_{aa'}({-} \iu \vec{\nabla}) \hat{\psi}_{a'}(\vec{x}) \\
	  	&\! + \frac{1}{2} \sum_{aa'} \int_{\vec{x} \vec{x}'} \hat{\psi}^{\dagger}_{a}(\vec{x}) \hat{\psi}^{\dagger}_{a'}(\vec{x}') V(\vec{x} - \vec{x}') \hat{\psi}_{a'}(\vec{x}') \hat{\psi}_{a}(\vec{x}) .
	\end{split}
\end{equation}
This shows that in a $\vmomentum \vec{\cdot} \vec{p}$ model the Coulomb interaction conserves the band index at the vertex only when it is written in the basis of orbitals at the $\Gamma$ point, or whichever $\vmomentum_{0}$ one chooses.
This has been derived from the most basic principles by starting with the Hamiltonian of Eq.~\eqref{eq:chMiscConc:generalInteractingHamiltonianMicroscopic} and using Bloch's theorem.
In our work, Eq.\ \eqref{eq:finalFormHeffFieldTheory} is used as the starting point after renaming $H_{0,\text{eff}}$ to $H_{0}$ again and assuming that it contains the Dirac-sea effects included in the first-principles calculation whose band structure we wish to replicate.
Note that we have quietly omitted these corrections in this appendix for simplicity, but they can be straightforwardly included by appropriately splitting the interaction, a point we return to in the next appendix.
Finally, we can expand the effective field operators in the eigenbasis of the $\vmomentum \vec{\cdot} \vec{p}$ Hamiltonian as
\begin{equation}
  \hat{\psi}_{a}(\vec{x}) = \frac{1}{\sqrt{\mathcal{A}}} \sum_{\alpha \vmomentum} \ec^{\iu \vmomentum \vec{\cdot} \vec{x}} \langle a \vert u^{\alpha}_{\vmomentum} \rangle \hat{\psi}_{\alpha \vmomentum} ,
\end{equation}
which in particular introduces the form factors of Eq.\ \eqref{eq:directInteraction} in the interaction term.

\section{Derivation of Eqs.\ \eqref{eq:genEVProblem}--\eqref{eq:meanField:gapParams}}
\label{app:derivationGenEVProblem}

\noindent
In this section we derive the generalized exciton eigenvalue problem at $T = 0$ given by Eq.\ \eqref{eq:genEVProblem}, as well as the generalized gap equation of Eq.\ \eqref{eq:meanField:gapParams}.
This is most conveniently done in the path-integral language.

\subsection{Path-integral setup and Hubbard-Stratonovich transformation}

\noindent
We set up the imaginary-time coherent-state path integral corresponding to the Hamiltonian of Eq.\ \eqref{eq:finalFormHeffFieldTheory} as
\begin{equation}
	\mathcal{Z} = \int \mathcal{D} \psi^{*} \mathcal{D} \psi \, \ec^{{-} S} ,
\end{equation}
where $\mathcal{Z}$ denotes the grand-canonical partition function and $S$ is the action.
The latter is written as $S = S_{0} + S_{\text{int}}$, with
\begin{subequations}
	\begin{align}
		S_{0} &= {-} \sum_{aa'} \int_{xx'} \psi^{*}_{a}(x) G_{0,aa'}^{-1}(x, x') \psi^{*}_{a'}(x')  , \\
		S_{\text{int}} &= \frac{1}{2} \sum_{aa'} \int_{xx'} \psi^{*}_{a}(x) \psi^{*}_{a'}(x') V(x - x') \psi^{*}_{a'}(x') \psi^{*}_{a}(x) .
	\end{align}
\end{subequations}
The fields $\psi$ and $\psi^{*}$ are Grassmann-valued and carry the relevant orbital and spin degrees of freedom.
To avoid clutter we have defined the combined position and imaginary-time variable $x = (\vec{x}, \tau)$, as well as the shorthand notation $\int_{x} = \int \mathrm{d}^{2} x \int_{0}^{\beta} \mathrm{d} \tau$.
The inverse Green's function reads
\begin{equation}
\label{eq:app:G0min1}
	G^{-1}_{0}(x, x') = {-} [\partial_{\tau} + H_{0}({-} \iu \vec{\nabla})] \delta(x - x') .
\end{equation}
Strictly speaking, $H_{0}$ here is not yet the $\vmomentum \vec{\cdot} \vec{p}$ Hamiltonian for \ce{AsO} or MTCO used in the main text, but rather a bare single-particle Hamiltonian which does not incorporate the effects due to the filled bands.
We elaborate on this point below.
The interaction is taken instantaneous in imaginary time, thus $V(x - x') = V(\vec{x} - \vec{x}') \delta(\tau - \tau')$.
For the purposes of this section we assume that the chemical potential is contained in $H_{0}$, thus we do not explicitly include it in Eq.\ \eqref{eq:app:G0min1}.

It is convenient to work in momentum-space, so we rewrite the interaction term as
\begin{align}
	&S_{\text{int}} = \frac{1}{2 \mathcal{A}} \sum_{aa'} \sum_{\vMomentum \vmomentum \vmomentum'} \int_{\tau} V(\vmomentum - \vmomentum') \\
	&\! \times \psi^{*}_{a, \vMomentum/2 + \vmomentum}(\tau) \psi^{*}_{a',  {-}\vMomentum/2 + \vmomentum'}(\tau) \psi_{a', {-} \vMomentum/2 + \vmomentum}(\tau) \psi_{a, \vMomentum/2 + \vmomentum'}(\tau) . \nonumber
\end{align}
To obtain the exciton eigenvalue equation we perform a Hubbard-Stratonovich transformation to the Fock field, which contains the interband excitations.
To this end we introduce a set of complex fields $\lambda^{aa'}_{\vMomentum \vmomentum}$ by multiplying $\mathcal{Z}$ by unity in the form $1 = \int \mathcal{D} \lambda \, \ec^{{-} S_{\lambda, 0}}$, where the free action for the $\lambda$ field reads
\begin{equation}
	S_{\lambda, 0} = \frac{1}{2 \mathcal{A}} \sum_{aa'} \sum_{\vMomentum \vmomentum \vmomentum'} \int_{\tau} [\lambda^{aa'}_{\vMomentum \vmomentum}(\tau)]^{*} V(\vmomentum - \vmomentum') \lambda^{aa'}_{\vMomentum \vmomentum'}(\tau) .
\end{equation}
We now shift the auxiliary $\lambda$ fields as $\lambda^{aa'}_{\vMomentum \vmomentum}(\tau) \rightarrow \lambda^{aa'}_{\vMomentum \vmomentum}(\tau) - \psi^{*}_{a', {-}\vMomentum/2 + \vmomentum}(\tau) \psi_{a, \vMomentum/2 + \vmomentum}(\tau)$ in their path integral.
This eliminates the quartic fermion term in the original action.
We note that, for this to be valid, the $\lambda$ fields cannot all be taken as independent but must rather satisfy the constraint $\lambda^{aa'}_{\vMomentum \vmomentum}(\tau) = [\lambda^{a'a}_{{-} \vMomentum, \vmomentum}(\tau)]^{*}$.
The shift also leads to a fermionic self-energy $\Sigma_{aa'}(\vmomentum, \tau; \vmomentum', \tau') = \delta(\tau - \tau') \Sigma_{aa'}(\vmomentum, \vmomentum', \tau)$, where
\begin{equation}
\label{eq:selfEnergyFock}
	\Sigma_{aa'}(\vmomentum, \vmomentum', \tau) = {-}\frac{1}{\mathcal{A}} \sum_{\vec{p}} V(\vmomentum - \vec{p}) \lambda^{aa'}_{\vmomentum - \vmomentum', \vec{p} - (\vmomentum - \vmomentum')/2}(\tau) .
\end{equation}
At this stage the action is quadratic in the electronic fields, which can be integrated out exactly.
We then obtain the effective action for the $\lambda$ field as
\begin{equation}
	S_{\lambda, \text{eff}} = S_{\lambda, 0} - \operatorname{Tr} \log ({-} G^{-1}) ,
\end{equation}
where $G^{-1} = G_{0}^{-1} - \Sigma$ is the inverse electronic Green's function with a Fock self-energy correction.

We now perform a fluctuation expansion for the $\lambda$ field by writing $\lambda = \langle \lambda \rangle + \tilde{\lambda}$.
By demanding that the linear term in the fluctuations vanish, we find the expectation value of $\lambda$ as
\begin{equation}
\label{eq:app:selfConsCond}
	\langle \lambda^{aa'}_{\vMomentum \vmomentum}(\tau) \rangle = G_{aa'}(\vMomentum/2 + \vmomentum, \tau; {-} \vMomentum/2 + \vmomentum, \tau^{+}) ,
\end{equation}
where $\tau^{+} = \tau + \iu 0^{+}$.

\subsection{Gap equation}

\noindent
Eq.\ \eqref{eq:app:selfConsCond} is a self-consistency condition for $\langle \lambda \rangle$ from where the gap equation can be obtained.
We now particularize to our case of interest in the main text, namely to static, zero-momentum configurations for $\lambda$ at the mean-field level.
In other words, we replace $\lambda^{aa'}_{\vec{K} \vec{k}}(\tau) \rightarrow \langle \lambda^{aa'}_{\vec{K} \vec{k}}(\tau) \rangle = \delta_{\vec{K}, \vec{0}} \langle \lambda^{aa'}_{\vec{k}}\rangle$ in the effective action.
Strictly speaking, the expectation value of the $\lambda$ field is always nonzero because there are filled bands in the system.
It will be convenient to work with respect to the zero-temperature Dirac sea of the uncorrelated TI phase, so define an appropriate density matrix as follows:
\begin{align}
\label{eq:HSTFullOrbBasis:rhoOrbBasis}
	\rho^{aa'}_{\vmomentum} &= \langle \lambda^{aa'}_{\vmomentum} \rangle - {\langle \lambda^{aa'}_{\vmomentum} \rangle}_{\mathrm{DS}} \\
	&= \frac{1}{\beta} \sum_{n} G_{aa'}(\vmomentum, \iu \omega_{n}) - \lim_{\beta \rightarrow \infty} \frac{1}{\beta} \sum_{n} G^{\mathrm{DS}}_{aa'}(\vmomentum, \iu \omega_{n}) . \nonumber
\end{align}
Here, $G$ is the Green's function of the system in the ground state and $G_{\mathrm{DS}}$ is the free electron Green's function with Dirac-sea corrections.
In particular, $\rho = 0$ in the TI phase at zero temperature.

From Eq.\ \eqref{eq:app:selfConsCond} we can easily derive the mean-field gap equation from the main text, which is obtained by setting all fluctuations to zero.
The self-energy that enters the dressed Green's function reads $\Sigma_{aa'}(\vec{k}) = \Delta^{aa'}_{\vmomentum} + \Sigma^{\mathrm{F}}_{aa'}(\vmomentum)$, where
\begin{subequations}
	\begin{align}
		\label{eq:HSTFullOrbBasis:DeltaOrbBasis}
		\Delta^{aa'}_{\vmomentum} &= {-} \frac{1}{\mathcal{A}} \sum_{\vmomentum'} V(\vmomentum - \vmomentum') \rho^{aa'}_{\vmomentum'} , \\
		\Sigma^{\mathrm{F}}_{aa'}(\vmomentum) &= {-} \frac{1}{\mathcal{A}} \sum_{\vmomentum'} V(\vmomentum - \vmomentum') {\langle \lambda^{aa'}_{\vmomentum'}\rangle}_{\mathrm{DS}} .
	\end{align}
\end{subequations}
The first expression corresponds to the definition of the gap parameters in the combined spin-orbital basis.
The second is a Fock contribution to the single-particle Hamiltonian.
The inverse Green's function then reads
\begin{equation}
	\label{eq:HSTFullOrbBasis:GDeltaInv}
	G^{-1}(\vmomentum, \iu \omega_{n}) = \iu \omega_{n} - H_{0}(\vmomentum) - \Sigma_{\mathrm{F}}(\vmomentum) - \Delta(\vmomentum) ,
\end{equation}
where $\Delta(\vmomentum)$ has components $\Delta^{aa'}_{\vmomentum}$.
The self-energy contribution is thus $\Sigma_{\mathrm{F}} + \Delta$.
It is important to realize that it is in fact $H_{0} + \Sigma_{\mathrm{F}}$ (rather than just the bare $H_{0}$) that corresponds to the $\vmomentum \vec{\cdot} \vec{p}$ Hamiltonian of interest, as it includes Dirac-sea corrections that give rise to the correct $GW$ band structure at zero temperature.
In the main text we have redefined this corrected Hamiltonian as $H_{0}$ for simplicity.

Eqs.\ \eqref{eq:HSTFullOrbBasis:rhoOrbBasis}, \eqref{eq:HSTFullOrbBasis:DeltaOrbBasis}, and \eqref{eq:HSTFullOrbBasis:GDeltaInv} form the self-consistent loop to be solved for a potentially nontrivial ground state with $\rho, \Delta \neq 0$.
If desired we can now transform all quantities to the band basis by writing
\begin{equation}
	\rho^{aa'}_{\vmomentum} = \sum_{\alpha \beta} \langle a \vert u^{\alpha}_{\vmomentum} \rangle \rho^{\alpha \beta}_{\vmomentum} \langle u^{\beta}_{\vmomentum} \vert a' \rangle
\end{equation}
and similarly for $\Delta$ and the Green's functions.
Here, $\vert u^{\alpha}_{\vmomentum} \rangle$ is an eigenstate of the (Dirac-sea corrected) single-particle Hamiltonian with eigenvalue $\xi^{\alpha}_{\vec{k}} = \epsilon^{\alpha}_{\vmomentum} - \mu$.
In particular, the Dirac sea contribution to the density matrix becomes very simple in the band basis:
\begin{equation}
	{\langle \lambda^{\alpha \beta}_{\vmomentum} \rangle}_{\mathrm{DS}} = \delta_{\alpha \beta} [1 - \Theta(\xi^{\alpha}_{\vec{k}})] ,
\end{equation}
which is one for the filled bands and zero for the unfilled bands.
Transforming the Green's function of Eq.\ \eqref{eq:HSTFullOrbBasis:GDeltaInv} to the band picture yields the mean-field Hamiltonian $\mathcal{H}_{\Delta}$ defined in the main text.
Finally, applying this change of basis to Eq.\ \eqref{eq:HSTFullOrbBasis:DeltaOrbBasis} and using the completeness relations of the single-particle eigenstates straightforwardly yields Eq.\ \eqref{eq:meanField:gapParams} in terms of the direct interaction matrix elements defined in Eq.\ \eqref{eq:directInteraction}.

\subsection{Fluctuations in the semiconductor phase}

\noindent
In the previous section we focused on a potentially nontrivial ground state ($\rho \neq 0$) by setting the fluctuations to zero.
In this section we consider the TI phase with $\rho = 0$ and derive the equation of motion for the fluctuations on top of this ground state.
These contain interband transitions associated with excitons and will be described by Eq.\ \eqref{eq:genEVProblem} at $T = 0$.

At the quadratic level, the action for the fluctuations on top of the Dirac-sea ground state is given by
\begin{equation}
	S^{(2)}_{\tilde{\lambda}, \text{eff}} = S_{\tilde{\lambda}, 0} + \frac{1}{2} \operatorname{Tr} G_{\mathrm{DS}} \tilde{\Sigma} G_{\mathrm{DS}} \tilde{\Sigma} ,
\end{equation}
where $\tilde{\Sigma}$ is the self-energy \eqref{eq:selfEnergyFock} with $\lambda$ replaced by the fluctuations $\tilde{\lambda}$.
Expanding the second term in momentum and frequency space we can write this explicitly as
	\begin{align}
	\label{eq:S2fluctExpl}
			&S^{(2)}_{\tilde{\lambda}, \text{eff}} = \frac{1}{2 \mathcal{A}} \sum [\tilde{\lambda}^{aa'}_{\vMomentum \vec{p} n}]^{*} V(\vec{p} - \vmomentum) \\
			&\times \bigg\{\delta_{ab} \delta_{a'b'} \delta_{\vmomentum \vmomentum'} + \frac{1}{\mathcal{A}} \Pi_{aa'bb'}(\vMomentum, \vmomentum, \iu \Omega_{n}) V(\vmomentum - \vmomentum')\bigg\} \tilde{\lambda}^{bb'}_{\vMomentum \vmomentum' n} , \nonumber
	\end{align}
	where the sum runs over all variables and $\Omega_{n}$ is a bosonic Matsubara frequency.
	This equation defines the inverse Green's function for the $\tilde{\lambda}$ field.
	The polarization bubble reads
	\begin{align}
		&\Pi_{aa'bb'}(\vMomentum, \vmomentum, \iu \Omega_{n}) \\
		&= \frac{1}{\mathcal{\beta}} \sum_{m} G^{\mathrm{DS}}_{ab}(\vMomentum/2 + \vmomentum, \iu \Omega_{n} + \iu \omega_{m}) G^{\mathrm{DS}}_{b'a'}({-}\vMomentum/2 + \vmomentum, \iu \omega_{m}) , \nonumber
	\end{align}
	where $\omega_{m}$ is a fermionic Matsubara frequency.
	This can be performed explicitly by expressing the Green's function in the band basis as
	\begin{equation}
	\label{eq:app:GDSBandExp}
		G^{\mathrm{DS}}_{aa'}(\vmomentum, \iu \omega_{m}) = \sum_{\alpha} \frac{\langle a \vert u^{\alpha}_{\vmomentum} \rangle \langle u^{\alpha}_{\vmomentum} \vert a' \rangle}{\iu \omega_{m} - \xi^{\alpha}_{\vec{k}}}
	\end{equation}
	and using the standard result
	\begin{equation}
	\label{eq:app:MatsubaraSum}
		\frac{1}{\mathcal{\beta}} \sum_{m} \frac{1}{\iu \Omega_{n} + \iu \omega_{m} - \xi_{1}} \frac{1}{\iu \omega_{m} - \xi_{2}} = \frac{N_{\mathrm{F}}(\xi_{1}) - N_{\mathrm{F}}(\xi_{2})}{\xi_{1} - \xi_{2} - \iu \Omega_{n}} .
	\end{equation}

	The equation of motion for the fluctuations corresponds to the nullspace of the inverse Green's function defined via Eq.\ \eqref{eq:S2fluctExpl} after performing the analytic continuation $\iu \Omega_{n} \rightarrow \Omega + \iu 0^{+}$.
	The resulting equation can be solved at each $\vMomentum$ due to the conservation of the total momentum, with the corresponding energy being denoted by $\Omega_{\vMomentum}$.
	We now use Eqs.\ \eqref{eq:app:GDSBandExp}--\eqref{eq:app:MatsubaraSum} and expand the $\tilde{\lambda}$ configuration satisfying the equation of motion in the band basis as
	\begin{equation}
		\tilde{\lambda}^{aa'}_{\vMomentum \vmomentum} = \sum_{\alpha \beta} \langle a \vert u^{\alpha}_{\vMomentum/2 + \vmomentum} \rangle \langle u^{\beta}_{{-}\vMomentum/2 + \vmomentum} \vert a' \rangle \varphi^{\alpha \beta}_{\vMomentum \vmomentum} .
	\end{equation}
	By projecting the equation of motion on an appropriate pair of single-particle states and multiplying through by the energy denominator of the second term we finally obtain the following coupled set of equations for the coefficients $\varphi$:
\begin{widetext}
	\begin{equation}
			(\xi^{\alpha}_{\vMomentum/2 + \vmomentum} - \xi^{\beta}_{{-}\vMomentum/2 + \vmomentum}) \varphi^{\alpha \beta}_{\vMomentum \vmomentum} + \frac{1}{\mathcal{A}} [N_{\mathrm{F}}(\xi^{\alpha}_{\vMomentum/2 + \vmomentum}) - N_{\mathrm{F}}(\xi^{\beta}_{{-}\vMomentum/2 + \vmomentum})] \sum_{\mathclap{\alpha' \beta' \vmomentum'}} V^{\mathrm{D}}_{\alpha \beta \alpha' \beta'}(\vMomentum, \vmomentum, \vmomentum') \varphi^{\alpha' \beta'}_{\vMomentum \vmomentum'} = \Omega_{\vMomentum} \varphi^{\alpha \beta}_{\vMomentum \vmomentum} .
	\end{equation}
\end{widetext}
	Here,
	\begin{equation}
		\begin{split}
			&V^{\mathrm{D}}_{\alpha \beta \alpha' \beta'}(\vMomentum, \vmomentum, \vmomentum') \\
			& = V(\vmomentum - \vmomentum') \langle u^{\alpha}_{\vMomentum/2 + \vmomentum} \vert u^{\alpha'}_{\vMomentum/2 + \vmomentum'} \rangle  \langle u^{\beta'}_{{-}\vMomentum/2 + \vmomentum'} \vert u^{\beta}_{{-}\vMomentum/2 + \vmomentum} \rangle
		\end{split}
	\end{equation}
	is a generalization of the direct matrix elements to nonzero total momentum $\vMomentum$.
	At zero temperature the equations acquire a particularly simple form.
	In this case, the difference of Fermi factors vanishes when the involved bands are both either empty or filled, and we obtain $\varphi^{cc'} = \varphi^{vv'} = 0$.
	Consequently, only interband transitions (described by $\varphi^{cv}$ and $\varphi^{vc}$) play a role in the low-temperature limit.
	Particularizing to the case of interest $\vMomentum = \vec{0}$ for simplicity (and thus omitting this label as in the main text), we obtain
\begin{widetext}
	\begin{subequations}
		\begin{align}
			\label{eq:fullEVPrEq1}
			(\xi^{c}_{\vmomentum} - \xi^{v}_{\vmomentum}) \varphi^{cv}_{\vmomentum} - \frac{1}{\mathcal{A}} \sum_{\mathclap{c' v' \vmomentum'}} V^{\mathrm{D}}_{c v c' v'}(\vmomentum, \vmomentum') \varphi^{c' v'}_{\vmomentum'} - \frac{1}{\mathcal{A}} \sum_{\mathclap{c' v' \vmomentum'}} V^{\mathrm{D}}_{c v v' c'}(\vmomentum, \vmomentum') \varphi^{v' c'}_{\vmomentum'} = \Omega \varphi^{c v}_{\vmomentum} , \\
			\label{eq:fullEVPrEq2}
			(\xi^{v}_{\vmomentum} - \xi^{c}_{\vmomentum}) \varphi^{vc}_{\vmomentum} + \frac{1}{\mathcal{A}} \sum_{\mathclap{c' v' \vmomentum'}} V^{\mathrm{D}}_{v c v' c'}(\vmomentum, \vmomentum') \varphi^{v' c'}_{\vmomentum'} + \frac{1}{\mathcal{A}} \sum_{\mathclap{c' v' \vmomentum'}} V^{\mathrm{D}}_{v c c' v'}(\vmomentum, \vmomentum') \varphi^{c' v'}_{\vmomentum'} = \Omega \varphi^{v c}_{\vmomentum} ,
		\end{align}
	\end{subequations}
\end{widetext}
	where we note that $\xi^{c}_{\vmomentum} - \xi^{v}_{\vmomentum} = \epsilon^{c}_{\vmomentum} - \epsilon^{v}_{\vmomentum}$ as the chemical potential drops out of this difference.
	Eqs.\ \eqref{eq:fullEVPrEq1}--\eqref{eq:fullEVPrEq2} can be rearranged into a matrix problem for the vector $(\varphi^{cv}, \varphi^{vc})$, denoted by $(\Phi, \bar{\Phi})$ in the main text.
	It is easy to verify that the interaction matrix elements satisfy the property
	\begin{equation}
		V^{\mathrm{D}}_{\alpha \beta \alpha' \beta'}(\vmomentum, \vmomentum') = V^{\mathrm{D}}_{\beta \alpha \beta' \alpha'}(\vmomentum, \vmomentum')^{*} .
	\end{equation}
	This immediately leads to the generalized eigenvalue problem of Eq.\ \eqref{eq:genEVProblem}.
	We see that, even at $T = 0$, the equation of motion for the fluctuations of the interband fields contain the $\mathrm{U}(1)$ breaking channels arising from the pair-production and annihilation processes sketched in Fig.\ \hyperref[fig:figIntro]{1b}.

	The above equations are correct at $\vMomentum = \vec{0}$ whenever the bare interaction only depends on distance and not on orbital character, as is the case here.
	When $\vMomentum \neq \vec{0}$, or even at $\vMomentum = \vec{0}$ with an orbital-dependent potential, the above derivation fails to account for the exchange interaction \cite{maisel2023single,wu2015exciton}.
	The latter arises from a Hartree-like field, and thus our pure Fock path-integral formalism does not take it into account.
	Nevertheless, it is straightforward to transform the equations of motion to the right form by replacing $V^{\mathrm{D}}$ with $V^{\mathrm{D}} - V^{\mathrm{X}}$ everywhere, where the exchange interaction is given by
	\begin{equation}
		\begin{split}
			&V^{\mathrm{X}}_{\alpha \beta \alpha' \beta'}(\vMomentum, \vmomentum, \vmomentum') \\
			&= V(\vMomentum) \langle u^{\alpha}_{\vMomentum/2 + \vmomentum} \vert u^{\beta}_{{-}\vMomentum/2 + \vmomentum} \rangle  \langle u^{\beta'}_{{-} \vMomentum/2 + \vmomentum'} \vert u^{\alpha'}_{\vMomentum/2 + \vmomentum'} \rangle .
		\end{split}
	\end{equation}

\section{Superfluid weight}

\noindent
The expression used to compute the superfluid weight of the excitonic insulator reads as follows (note that we consider a fixed spin subspace and omit the label):
\begin{equation}
\label{eq:DsMuNuConv}
	\begin{split}
		D_{s}^{\mu \nu} &= \frac{1}{\mathcal{A}} \sum_{\alpha \beta \vmomentum} \frac{N_{\mathrm{F}}(\omega^{\alpha}_{\vmomentum} - \mu) - N_{\mathrm{F}}(\omega^{\beta}_{\vmomentum} - \mu)}{\omega^{\alpha}_{\vmomentum} - \omega^{\beta}_{\vmomentum}} \\
		&\times \Big(\langle \psi^{\alpha}_{\vmomentum} \vert \gamma_{z} \partial_{\mu} H_{0}^{(\mathrm{d})}(\vmomentum) \vert \psi^{\beta}_{\vmomentum} \rangle \langle \psi^{\beta}_{\vmomentum} \vert \gamma_{z} \partial_{\nu} H_{0}^{(\mathrm{d})}(\vmomentum) \vert \psi^{\alpha}_{\vmomentum} \rangle \\
		&\hspace{1.5mm} - \langle \psi^{\alpha}_{\vmomentum} \vert \partial_{\mu} \mathcal{H}^{\Delta}(\vmomentum) \vert \psi^{\beta}_{\vmomentum} \rangle \langle \psi^{\beta}_{\vmomentum} \vert \partial_{\nu} H_{0}^{(\mathrm{d})}(\vmomentum) \vert \psi^{\alpha}_{\vmomentum} \rangle \Big) .
	\end{split}
\end{equation}
Here, $\omega^{\alpha}_{\vmomentum}$ and $\vert \psi^{\alpha}_{\vmomentum} \rangle$ are the eigenvalues and eigenstates of $\mathcal{H}^{\Delta}(\vmomentum)$, the mean-field Hamiltonian, $H^{(\mathrm{d})}_{0}(\vmomentum) = \operatorname{diag}(\epsilon^{c}_{\vmomentum}, \epsilon^{v}_{\vmomentum})$ is the diagonalized single-particle Hamiltonian, and $\gamma_{z}$ is the Pauli matrix acting on the conduction-valence isospin degree of freedom.

We note that the $\gamma_{z}$ operator in this context appears in the opposite term with respect to the formulas in Refs.\ \cite{kitamura2022quantum,kitamura2022superconductivity}, where it is found in front of the $H^{(\mathrm{d})}_{0}$ term in the last line.
This is because these works consider a superconductor, where the second diagonal block in the Bogoliubov-de Gennes Hamiltonian includes a minus sign.
This block plays the role of our valence sector \emph{without} this minus sign.
Following the derivation of $D_{s}$ adapted to our EI case results in Eq.\ \eqref{eq:DsMuNuConv}.
We also note that the expression used in Ref.\ \cite{yang2024spin} is valid only for in for momentum-independent gap parameter, which is not the case at hand.
For the conventional part of the superfluid weight in the case of $\vmomentum$-dependent $\Delta$'s, Eq.\ \eqref{eq:DsMuNuConv} must be used.

\bibliography{bib}

\end{document}